\documentclass[fleqn,usenatbib]{mnras}

\newcommand{\etal}{\mbox{et\ al.\ }}

\usepackage{newtxtext,newtxmath}
\usepackage[T1]{fontenc}
\usepackage{aecompl}   
\usepackage{amsmath}    
\usepackage{amssymb}
\usepackage{graphicx,changepage}
\usepackage{rotating}
\usepackage{url}
\usepackage{natbib}
\usepackage{booktabs}
\usepackage{lscape}
\usepackage{longtable}

\title[Investigating \textit{INTEGRAL} GRBs]
{Investigating the nature of \textit{INTEGRAL} Gamma-ray Bursts and sub-threshold triggers with \textit{Swift} follow-up}

\author[A.B. Higgins \etal]
{A. B. Higgins,$^{1}$\thanks{Email: abh13@le.ac.uk} R. L. C. Starling,$^1$ D. G{\"o}tz,$^2$ S. Mereghetti,$^3$ K. Wiersema,$^1$ \and T. Maccarone,$^4$ J. P. Osborne,$^1$ N. R. Tanvir,$^1$ P. T. O'Brien,$^1$ A. J. Bird,$^5$ \and A. Rowlinson$^{6,7}$  and N. Gehrels$^8$\\
$^1$Department of Physics and Astronomy, University of Leicester, University Road, Leicester LE1 7RH, UK.\\
$^2$AIM-CEA/DRF/Irfu/Service d'Astrophysique, Orme des Merisiers, 91191 Gif-sur-Yvette, France.\\
$^3$INAF, IASF-Milano, via E.Bassini 15, 20133 Milano, Italy.\\
$^4$Department of Physics and Astronomy, Texas Tech University, Box 41051, Lubbock, TX 79409, USA.\\
$^5$School of Physics and Astronomy, University of Southampton, Southampton SO17 1BJ, UK. \\
$^6$Netherlands Institute for Radio Astronomy (ASTRON), PO Box 2, 7990 AA Dwingeloo, The Netherlands.\\
$^7$Anton Pannekoek Institute, University of Amsterdam, Postbus 94249, 1090 GE, Amsterdam, The Netherlands.\\
$^8$NASA Goddard Space Flight Center, Greenbelt, MD 20771, USA.}

\begin{document}
\date{Accepted . Received ; in original form .}

\pagerange{\pageref{firstpage}--\pageref{lastpage}} \pubyear{}

\maketitle

\label{firstpage}


\begin{abstract}
We explore the potential of \textit{INTEGRAL} to improve our understanding of the low fluence regime for explosive transients, such as GRBs.
We probe the nature of the so-called "WEAK" \textit{INTEGRAL} triggers, when the gamma-ray instruments record intensity spikes that are below the usual STRONG significance thresholds. In a targeted \textit{Swift} follow-up campaign, we observed 15 WEAK triggers. We find six of these can be classified as GRBs. This includes GRB\,150305A, a GRB discovered from our campaign alone. We also identified a source coincident with one trigger, IGRW\,151019, as a candidate AGN. We show that real events such as GRBs exist within the IBAS WEAK trigger population.
A comparison of the fluence distributions of the full \textit{INTEGRAL} IBAS and \textit{Swift} BAT GRB samples showed that the two are similar. We also find correlations between the prompt gamma-ray and X-ray properties of the two samples, supporting previous investigations. We find that both satellites reach similar, low fluence levels regularly, although \textit{Swift} is more sensitive to short, low fluence GRBs.
\end{abstract}


\section{Introduction} \label{sec:intro}
GRBs are among the most luminous events in the universe, releasing energies $>10^{51}$ erg typically in time periods of seconds \citep{Gehrels2012}. During these events a huge amount of gravitational energy is released from a central engine which leads to the formation of jets where particles are accelerated to ultra-relativistic speeds \citep{Woosley2006b}. Internal shocks within the jet produce the high energy prompt gamma-ray emission we first observe \citep{Gehrels2012,Piran2003}. The jet then shocks with the surrounding medium producing broad-band afterglow emission \citep{Meszaros1997,Wijers1997}. Classically, GRBs are split into two sub-groups based on their T$_{90}$ - the duration over which 90 per cent of the gamma-ray flux is received \citep{Kouveliotou1993}. The two groups are short GRBs where T$_{90} < 2$ s and long GRBs where T$_{90} > 2$ s, linked with two different progenitor models.

GRBs span a large range of isotropic equivalent luminosities: $10^{45} \leq {\rm L_{ISO}} \leq 10^{54}$ erg s$^{-1}$. Investigations into their luminosity function and formation rates coupled with observations of several local GRBs \citep{Sazonov2004,Soderberg2004} have suggested that there should be a large number of low-luminosity GRBs \citep{Daigne2007,Liang2007,Pescalli2016}. There are further suggestions that these could exist as a separate local population \citep{Norris2002,Norris2005,Chapman2007,Liang2007}. 
We use two currently active GRB detecting missions, \textit{Swift} \citep{Gehrels2004} and The INTErnational Gamma-Ray Astrophysics Laboratory (\citealt{Winkler2003}; INTEGRAL), to look at potentially faint GRBs.

\textit{INTEGRAL} carries two gamma-ray instruments, IBIS \citep{Ubertini2003} and SPI \citep{Vedrenne2003}. Alerts for GRBs and other transient sources are communicated with low latency by the \textit{INTEGRAL} Burst Alert System, IBAS\footnote{\label{IBAS} http://ibas.iasf-milano.inaf.it/} \citep{Mereghetti2003} discussed in more detail in section \ref{sec:IBASalerts}. Since the launch in 2002 \textit{INTEGRAL} has detected over 900 soft gamma-ray sources\footnote{\label{IntegralWeb} http://www.isdc.unige.ch/integral/} \citep{Bird2016} and has localised 114 GRBs (numbers correct as of 2016 July 1). \textit{INTEGRAL} has made some important discoveries regarding GRBs, reviewed in \cite{Gotz2012} including investigations utilising IBIS and SPIs capability to perform spectral analysis on the \textit{INTEGRAL} sample of GRBs \citep{Vianello2009,Bosnjak2014}. Furthermore, \cite{Foley2008} suggested that \textit{INTEGRAL} may be capable of detecting the local, low-luminosity GRB populations.

In the fully coded field of view (FOV), i.e. the central $9^{\circ}\times 9^{\circ}$, the \textit{INTEGRAL} IBIS instrument is more sensitive than the Burst Alert Telescope (BAT) \citep{Barthelmy2004} on board \textit{Swift}, despite its smaller effective area (2600 cm$^2$ compared to 5200 cm$^2$). This is due to the fact that, at the energies we are interested in ($15-200$ keV), the background is dominated by the Cosmic X-ray diffuse emission, which is proportional to the FOV (a factor of about ten smaller for IBIS than for BAT). Therefore \textit{INTEGRAL} should be able to reach lower peak flux limits, especially for GRBs with hard spectra where peak energies $>50$ keV \citep{Bosnjak2014}. However, since \textit{INTEGRAL} spends a large fraction of its observing time observing at low Galactic latitudes, its sensitivity is reduced by the additional background caused by bright Galactic sources and hard X-ray Galactic diffuse emission. It is only since the \textit{INTEGRAL} sub-threshold trigger campaign began (see section \ref{sec:IBASalerts}) that lower sensitivities have been routinely accessible through WEAK alerts.

\textit{Swift} has two additional instruments, the X-Ray Telescope (XRT) \citep{Burrows2005a} and the Ultraviolet/Optical Telescope (UVOT) \citep{Roming2005}, and has the ability to slew  towards a BAT-detected burst or pre-selected target. Therefore it can complement \textit{INTEGRAL} with rapid multi-wavelength, follow-up measurements. Using observations from both satellites we expect to uncover both the temporal behaviour and energetics of both the WEAK alerts and \textit{INTEGRAL} GRB sample and characterise their properties.

We start by discussing the \textit{INTEGRAL} Burst Alert System (IBAS) in more detail and describe our chosen WEAK triggers in section \ref{sec:IBASalerts}. Our \textit{Swift} follow-up analysis is discussed in section \ref{sec:swiftanalysis}. These are then analysed in conjunction with the total IBAS GRB sample in section \ref{sec:IBASsamps} with some comparisons to the \textit{Swift} GRB population. We conclude with our summary in section \ref{sec:concl}. 

\section{INTEGRAL IBAS Alerts} \label{sec:IBASalerts}
\textit{INTEGRAL} was designed as a general purpose gamma-ray observatory, not specifically optimized for the study of GRBs. However, its good imaging capabilities over a field of view of $\approx 30\times 30^{\circ}$ ($9\times 9^{\circ}$ fully coded and $19\times 19^{\circ}$ half coded) and the continuous telemetry downlink (due to its high elliptical orbit with a period of 3 days) made it possible to search and localize GRBs on the ground in near real time. This is done with the \textit{INTEGRAL} Burst Alert System, IBAS \citep{Mereghetti2003}, software running at the \textit{INTEGRAL} Science Data Centre, ISDC \citep{Courvoisier2003} since the launch of \textit{INTEGRAL} in October 2002.

No GRB triggering algorithm is implemented on board the satellite. The data reach the ISDC typically within 20 seconds after they have been collected and are immediately fed into the IBAS software which exploits several burst detection programs in parallel. When a burst (or any other new transient source) is detected inside the field of view of the IBIS instrument, its coordinates are automatically distributed via the Internet by means of Alert Packets based on the User Datagram Protocol (UDP). Their coordinates derived by IBAS have a mean with $1\sigma$ uncertainty of 2.1($\pm0.5$) arcmin. 

IBAS also searches for GRBs detected in the Anti-Coincidence Shield (ACS) of the SPI instrument, which provides a good sensitivity over nearly the whole sky, but without localisation and spectral information \citep{vonKienlin2003}. The ACS lightcurves are used for GRB localizations by triangulation with other satellites of the IPN network \citep{Cline1999}. In this investigation we will not discuss SPI ACS results. 

The search for GRBs in the IBIS data uses two different kinds of programs: rate monitor and image monitor programs. Rate monitors look for excesses in the light curve of the whole detection plane, while image monitors search for excesses in the deconvolved images. Both use data from ISGRI \citep{Lebrun2003}, the lower energy detector of IBIS, which provides photon by photon data in the energy range 15 keV - 1 MeV. Several instances of the rate and image monitors run in parallel using different settings for integration time scales and energy range. When one (or typically more) of the monitor programs triggers, an imaging analysis is performed on an optimally selected time interval in order to confirm the source presence and derive its significance.

Two significance threshold levels, labelled STRONG and WEAK, have been implemented in IBAS for what concerns the distribution of Alert Packets. The positions of new sources with significance above the STRONG threshold are immediately distributed with Alert Packets. These positions automatically derived by the IBAS software can be later refined by interactive analysis. Until 2011, Alert Packets for sources with significance above the WEAK threshold and below the STRONG were distributed in real time only to members of the IBAS Team, who, after interactive analysis could in some cases confirm the presence of a GRB and distribute its coordinates. However, in the majority of the cases it was not possible, based on the \textit{INTEGRAL} data alone, to confirm the real astrophysical nature of these low significance events. Since 2011 January 26, all the Alert Packets corresponding to detections above the WEAK threshold have been automatically distributed in real time to the external users who wish to receive them.

Among the 114 confirmed GRBs detected by IBAS, 17 have been detected as sub-threshold WEAK alerts and 54 were observed with \textit{Swift}, either through an independent autonomous BAT trigger and subsequent follow-up, or via ToO follow-up that was uploaded at a later time, and have available XRT data.

\subsection{Selection of WEAK alerts and follow-up} \label{sec:swifttoo}
There have been 402 \textit{INTEGRAL} WEAK triggers, below $8\sigma$ significance, before 2016 July 1; six of which were promoted to STRONG triggers and were later confirmed as GRBs. Out of the other 396 we analysed 15 WEAK triggers. They consisted of:

\begin{itemize}
\item{11 triggers that did not have prompt \textit{Swift} slews and were target of opportunity observations from our campaign. We named them IGRWYYMMDD prior to source-type identification, broadly following the GRB naming convention, see table \ref{tab:toos}. These are termed as "our chosen ToOs".}
\item{Two other WEAK \textit{INTEGRAL} triggers with ToOs requested elsewhere and had XRT data, but were not related to our 11 chosen triggers, were analysed. These are termed as "candidate GRBs".}
\item{Two WEAK triggers that also triggered BAT and had XRT data were also analysed. These are also termed as "candidate GRBs".}
\end{itemize} 
    
Candidate triggers for our \textit{Swift} ToO follow-up were selected with the requirement that at least one of the following criteria were met. Firstly, triggers were chosen to be close to the $8\sigma$ STRONG threshold (our lowest was $6.7\sigma$). This was to increase the chance of the trigger representing a real GRB. Trigger positions were also checked for high Galactic extinction and close proximity to nearby catalogued X-ray sources. Finally, triggers were generally only followed up if the trigger time coincided with the working hours of the on-call member of the \textit{Swift} team. The criteria described above were not stringently adhered to for all triggers. We cannot claim that these triggers form a uniform or complete sample and biases towards high significance and lower Galactic column density are present. This was a pilot campaign aimed at determine whether real transient events exist among the WEAK trigger population and we stress that we do not make conclusions for the entire WEAK trigger population.

\renewcommand{\arraystretch}{0}
\begin{table*}
    \centering
	\begin{tabular}{|c|c|c|c|c|c|c|c|c|c|c|}
    \hline
  \multicolumn{1}{|p{1.5cm}|}{\centering ToO \\ Name}
& \multicolumn{1}{|p{1.5cm}|}{\centering \textit{INTEGRAL} \\ Trigger No.} 
& \multicolumn{1}{|p{1.7cm}|}{\centering IBAS detection Significance ($\sigma$)}
& \multicolumn{1}{|p{1.5cm}|}{\centering RA \\ (Deg) \\ (J2000)}
& \multicolumn{1}{|p{1.5cm}|}{\centering Dec \\ (Deg) \\ (J2000)}
& \multicolumn{1}{|p{1.5cm}|}{\centering Localisation \\ Error \\ (arcmin)} \\ \hline
IGRW\,160610 & 7488/0 & 6.7 & 359.90 & 61.57 & 3.8 \\ \hline
IGRW\,151019 & 7277/0 & 7.0 & 292.82 & 31.14 & 3.5 \\ \hline
IGRW\,150903 & 7231/0 & 6.7 & 239.17 & -33.81 & 3.6 \\ \hline
IGRW\,150610 & 7005/0 & 7.1 & 178.32 & 16.03 & 4.8 \\ \hline
IGRW\,150305 & 6905/0 & 7.6 & 269.79 & -42.62 & 3.4 \\ \hline
IGRW\,140219 & 6467/0 & 6.7 & 204.10 & -45.06 & 3.6 \\ \hline
IGRW\,130904 & 6931/0 & 6.7 & 256.88 & -32.01 & 3.6 \\ \hline
IGRW\,110718 & 6323/0 & 6.8 & 256.78 & 40.05 & 3.6 \\ \hline
IGRW\,110608 & 6297/0 & 6.8 & 315.28 & 32.041 & 3.6 \\ \hline
IGRW\,110428 & 6169/0 & 7.2 & 320.27 & -33.96 & 3.5 \\ \hline
IGRW\,110112 & 6127/0 & 7.4 & 10.56 & 64.41 & 2.6 \\ \hline \\ \hline
IGRW\,150831 & 7228/0 & 7.3 & 220.98 & -25.65 & 3.4 \\ \hline
IGRW\,121212 & 6720/0 & 7.9 & 177.90 & 78.00 & 3.3 \\ \hline
IGRW\,100909 & 6060/0 & 7.7 & 73.95 & 54.65 & 2.0 \\ \hline
IGRW\,091111 & - & 7.2 & 137.81 & -45.91 & 2.9 \\ \hline
	\end{tabular}
 	\caption[ToO Table]{Table containing the properties of the 15 WEAK triggers. The four triggers at the bottom are the four candidate GRBs with previous XRT observations and were not part of our selected ToOs. The Trigger No., significance ($\sigma$), RA, Dec and localisation error (90 per cent confidence) were all taken from IBAS.}
 	\label{tab:toos}
\end{table*}  

\section{\textit{Swift} Analysis} \label{sec:swiftanalysis}
The XRT and UVOT data from the 15 WEAK triggers with follow-up \textit{Swift} observations discussed in section \ref{sec:swifttoo} were analysed to determine the nature of the WEAK trigger events. The data were made available by the UK \textit{Swift} Science Data Centre (UKSSDC) \citep{Evans2007,Evans2009}.

Cleaned event files for our 11 ToOs were produced using the \textit{Swift} XRT pipeline tool (v0.13.2). For the other four candidate GRBs we used the existing XRT products made available by the UKSSDC. For each ToO a search for any sources with a probability of being due to statistical fluctuations $< 0.3$ per cent (equivalent to $3\sigma$) within the \textit{INTEGRAL} error region (90 per cent confidence) was conducted using the sky image file. Source counts were derived from 30 arcsec radius regions centred on any detected X-ray source coordinates. Upper limits on non-detections were also obtained using Bayesian analysis described in \cite{Kraft1991}.

\begin{figure*}
\includegraphics[width=\textwidth]{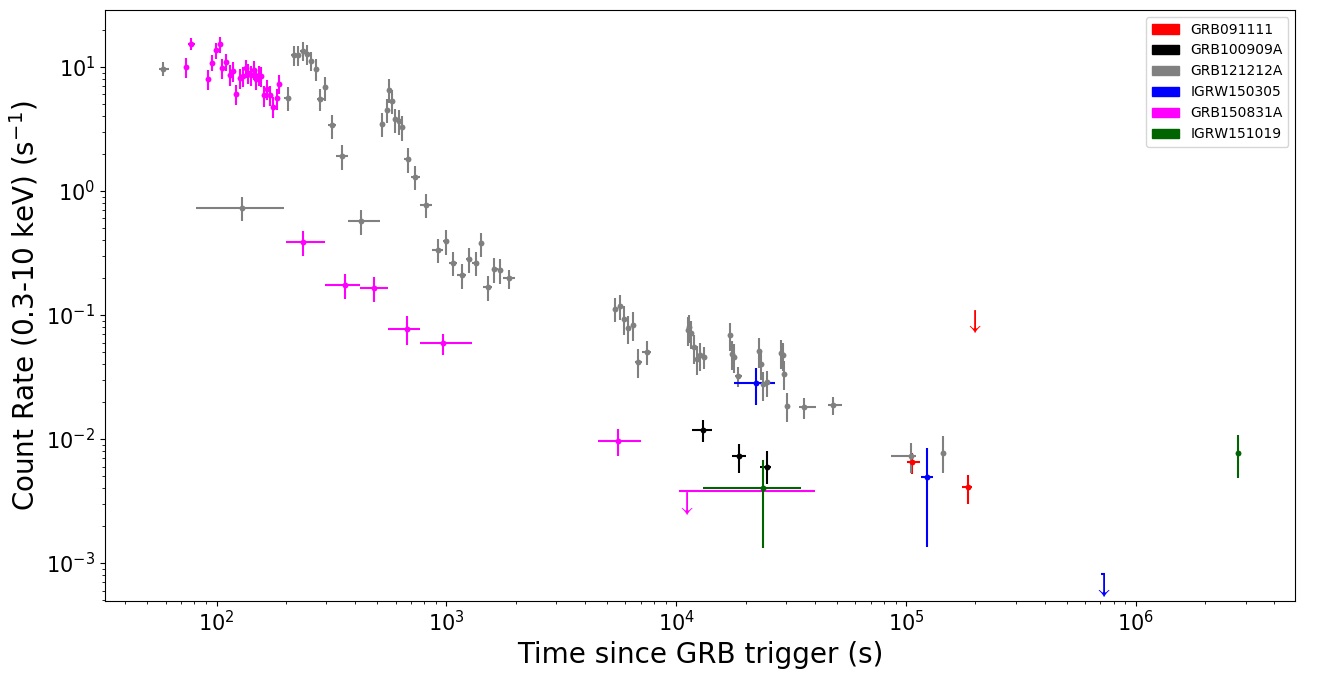}
\caption{X-ray afterglows of the six WEAK \textit{INTEGRAL} sources that were detected by \textit{Swift} from our 15 WEAK triggers.} 
\label{fig:weakdetplot} 
\end{figure*}

\begin{figure}
\includegraphics[width=\linewidth]{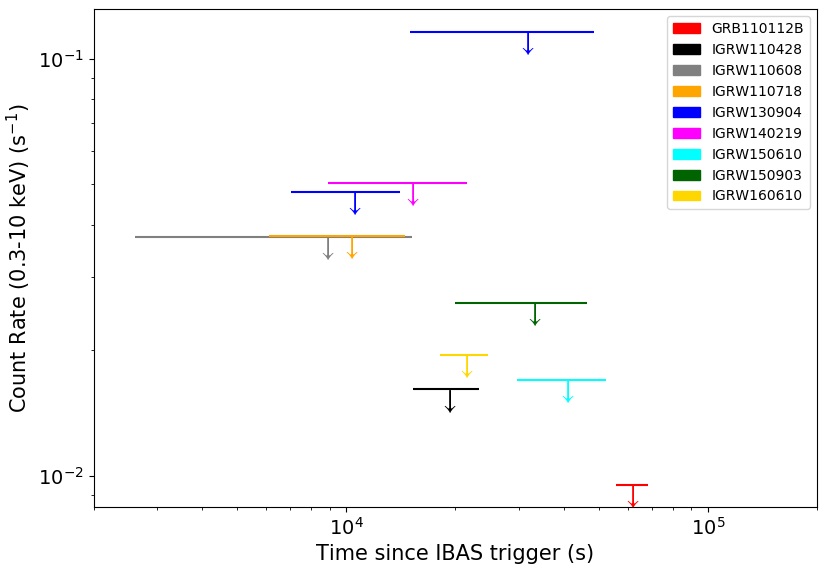}
\caption{XRT $3\sigma$ bayesian upper limits of the 9 non-detections out of our 15 selected WEAK triggers.} 
\label{fig:weaknondetplot}  
\end{figure}

If a source was detected with the \textit{Swift} XRT a further ToO observation was requested to identify whether the source was fading and thus could be confirmed as a GRB. If the source was detected again, and confirmed to be fading, a third observation was requested at a later date to check if the source had faded further. All positive detection coordinates were cross-referenced with the astrophysics catalogue database Vizier\footnote{\label{Vizier} http://vizier.u-strasbg.fr/viz-bin/VizieR} \citep{Ochsenbein2000} to identify any existing sources that could account for the X-ray emission. We obtained the following results:

\begin{itemize}
\item{For six of the 15 WEAK triggers, comprising of two of our chosen ToOs and the four candidate GRBs, we had a detection with the XRT. The \textit{Swift} XRT properties of these events, along with the non-detections, can be found in table \ref{tab:xrtobs}.}
\item{Subsequent observations found that five of these were fading X-ray sources, typical of a GRB afterglow \citep{Costa1997,OBrien2006} (see figure \ref{fig:weakdetplot}). The exception was IGRW\,151019 (discussed in section \ref{sec:IGRW151019results}).  The XRT non-detection upper limits can be seen in figure \ref{fig:weaknondetplot}.}
\item{All 6 positive X-ray detections had no previously catalogued X-ray sources within 2 arcmin at the time of the observations.}
\end{itemize}

\begin{figure}
\includegraphics[width=\linewidth]{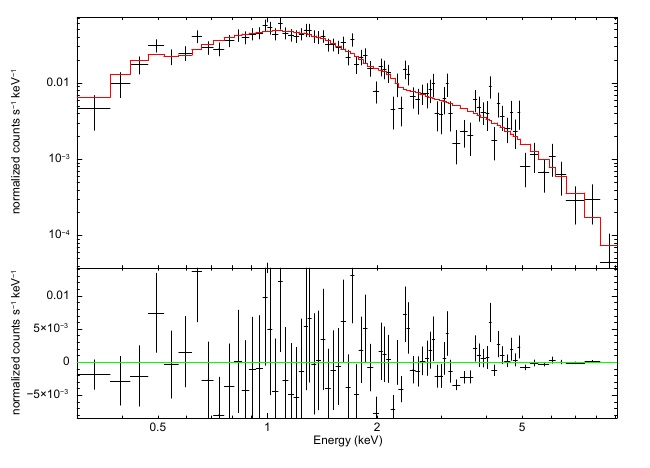}
\caption{XRT spectrum of GRB\,121212A with the best fitting absorbed power law model produced using {\sc xspec} (red). The fit parameters can be seen in table \ref{tab:weakspecproperties}.} 
\label{fig:GRB121212Aspec}  
\end{figure}

Two of the candidate GRBs, GRB\,121212A and GRB\,150831A, were relatively well observed by the XRT ($> 10$ data points) compared to the other WEAK trigger XRT sources as they also triggered BAT. These were further analysed to obtain both a spectral fit and X-ray afterglow decay slope. The results can be seen in table \ref{tab:weakspecproperties} and figure \ref{fig:GRB121212Aspec} shows the spectrum for GRB\,121212A. GRB\,150831A has a T$_{90}\approx 2$ s - classifying it as a short GRB. Although ToO IGRW\,110112 was an XRT non-detection at $6.2(\pm0.6)\times 10^{4}$ s after the IBAS trigger, its initial gamma-ray trigger was seen simultaneously by Fermi GBM \citep{Connaughton2011} and so was classified as a GRB. IGRW\,110608, one of the non-detections appeared to have an unusually high and irregular X-ray background compared to the other ToOs. This may have reduced our chances of getting a detection.

\renewcommand{\arraystretch}{0.5}
\begin{table*}
    \centering
	\begin{tabular}{|c|c|c|c|c|c|c|c|c|c|c|}
    \hline
  \multicolumn{1}{|p{1.4cm}|}{\centering ToO \\ Name}
& \multicolumn{1}{|p{1.0cm}|}{\centering \textit{Swift} \\ Obs ID} 
& \multicolumn{1}{|p{1.6cm}|}{\centering XRT \\ Position \\ Error \\ (arcsec)}
& \multicolumn{1}{|p{0.8cm}|}{\centering RA \\ (Deg) \\ (J2000)}
& \multicolumn{1}{|p{0.8cm}|}{\centering Dec \\ (Deg) \\ (J2000)}
& \multicolumn{1}{|p{1.0cm}|}{\centering $T_{\rm START}$ \\ (s)}
& \multicolumn{1}{|p{1.0cm}|}{\centering $T_{\rm STOP}$ \\ (s)}
& \multicolumn{1}{|p{0.8cm}|}{\centering $T_{\rm EXP}$ \\ (s)}
 \\ \hline
IGRW\,151019 & 20558 & 2.5 & 292.7836 & 31.1319 & 9600 & 2810017 & 14938 \\ \hline
GRB\,150831A & 653838 & 1.6 & 221.0243 & -25.6351 & 82 & 38575 & 11828 \\ \hline
IGRW\,150305A & 33663 & 3.5 & 269.7606 & -42.6638 & 17838 & 735268 & 908 \\ \hline
GRB\,121212A & 541371 & 1.4 & 177.7923 & 78.0371 & 60 & 145420 & 22940 \\ \hline
GRB\,100909A & 20147 & 3.3 & 73.9488 & 54.6579 & 11693 & 25787 & 7720 \\ \hline
GRB\,091111 & 20120 & 7.7 & 137.8233 & -45.9253 & 100360 & 197466 & 10386 \\ \hline 
	\end{tabular}
 	\caption[XRT Detection Table]{\textit{Swift} XRT ToO observations and candidate counterpart source detections. XRT position (90 per cent confidence), RA and Dec were taken from UKSSDC. $T_{\rm START}$ refers to the time elapsed between the GRB/trigger occurring and the time when the \textit{Swift} observation began and $T_{\rm STOP}$ refers to the time elapsed between the GRB occurring and the time when the final \textit{Swift} observation finished. $T_{\rm EXP}$ is the total XRT exposure time.}
 	\label{tab:xrtobs}
\end{table*}

\renewcommand{\arraystretch}{0.5}
\begin{table*}
\centering
	\begin{tabular}{|l|c|c|c|c|c|}
    \hline
  \multicolumn{1}{|p{2.0cm}|}{\centering ToO Name}
& \multicolumn{1}{|p{1.8cm}|}{\centering $N_{\rm H}$(Gal) ($10^{20}$ cm$^{-2}$)} 
& \multicolumn{1}{|p{1.8cm}|}{\centering $N_{\rm H}$(Int) ($10^{20}$ cm$^{-2}$)}
& \multicolumn{1}{|p{1.0cm}|}{\centering $\Gamma$}
& \multicolumn{1}{|p{1.1cm}|}{\centering C-Stat (dof)}
& \multicolumn{1}{|p{1.0cm}|}{\centering $\alpha$}
  \\ \hline
GRB\,121212A & 4.48 & $21^{+0.5}_{-0.4}$ & $2.24^{+0.14}_{-0.13}$ & 341 (369) &$-0.71^{+0.03}_{-0.03}$ \\ \hline
GRB\,150831A (WT) & 11.4 & $0^{+80.0}_{-0}$ & $1.15^{+0.18}_{-0.1}$ & 322 (404) & $-2.67^{+0.22}_{-0.22}$  \\ \hline
GRB\,150831A (PC) & 11.4 & $0^{+18.0}_{-0}$ & $1.53^{+0.28}_{-0.29}$ & 99 (93) & $-2.67^{+0.22}_{-0.22}$  \\ \hline
	\end{tabular}
 	\caption[WEAK Spectral Properties]{Table containing the X-ray spectral and afterglow light curve properties of WEAK GRBs with $>10$ binned data points. $N_{\rm H}$(Gal) is the fixed Galactic absorption column density and $N_{\rm H}$(Int) is the excess absorption. Spectral analysis was performed using {\sc xspec} and fitting an absorbed power law where $\Gamma$ is the photon index. The X-ray decay slopes were calculated using non-linear least squares fitting with various broken power law models. For each case a simple non-broken power law provided the best fit. All errors given at 90 per cent confidence level apart from the X-ray decay slopes - they are given at $1\sigma$.}
 	\label{tab:weakspecproperties}
\end{table*} 

\begin{figure}
\includegraphics[width=\linewidth]{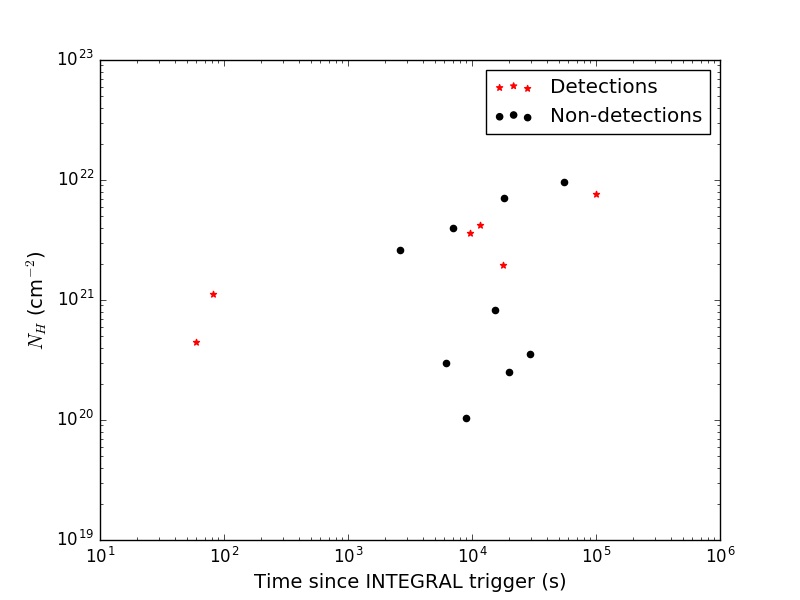}
\caption{Plot of $T_{\rm START}$ against Galactic $N_{\rm H}$ of all of our 15 WEAK triggers including both XRT detections and non-detections.}
\label{fig:Tstart_NH}
\end{figure}

In figure \ref{fig:Tstart_NH} we plot the time from the GRB to the start of the XRT observation ($T_{\rm START}$) against the weighted mean Galactic column density, $N_{\rm H}$(Gal) for both detections and non-detections. As the X-ray emission decays over time a later observation may result in a non-detection of an X-ray source that had been present at an earlier time. Additionally, high Galactic column density may reduce the chance of achieving a detection. The values for $N_{\rm H}$(Gal) were calculated using the method described in \cite{Willingale2013}. Two sources observed less than 100 s after the initial trigger were both detected. Of the other 13 sources observed at later times after the triggers four were detected and nine were not. The column density and time since the trigger values for these detections and non-detections were similar and from our observations we saw that the column density (up to $\approx 10^{22}$ cm$^{-2}$) and $T_{\rm START}$ (up to $\approx 70000$ s) had no significant impact as to whether a WEAK trigger would be detected by the XRT. However, we only observed 15 sources and two sources observed within 100 s of the trigger were both detected so observing sources as promptly as possible would aid in detecting any potential afterglows.

For the six X-ray detections we analysed the UVOT data to determine if any UV/optical sources were present. The data comprised one or more UVOT filters each with a number of separate images. Multiple images were aligned and summed up to create one image per exposure for each filter. If there were multiple exposures over all observations of the ToO these were also additionally summed together to create one image with the total exposure over all observations. It must be noted that the number of filters used during each ToO exposure was dependent upon those already designated to be used by \textit{Swift} on the date of the observation (the image binning could vary per exposure so only $1\times 1$ binned images were used during the investigation - see \textit{Swift} UVOT Online Manual).

To find the magnitude of a possible UVOT source, or upper limit on any non-detections, the \textit{Swift} tool {\sc uvotsource} was used with a significance of $3\sigma$ to distinguish between a possible source and non-detection upper limit \citep{Breeveld2010}. From our UVOT analysis we found:

\begin{itemize}
\item{An optical source was marginally detected ($<5\sigma$) with the UVOT white filter coincident with the XRT position of GRB\,121212A. A Vizier search of the source position revealed no reported optical source. The UVOT position of the GRB\,121212A optical source was RA, Dec (J2000) 177.79341, 78.03780 deg with a $1\sigma$ positional error of 0.48 arcsec.}
\item{A marginal detection was also registered in the v, b and u bands for GRB\,091111. Further analysis revealed that the event occurred within 30 arcsec of the centre of a very bright, saturated source which may have affected the background region near GRB\,091111 resulting in a false detection.}
\item{A UV source detected with the m2 filter ($9.2\sigma$) was present very close to the 90 per cent XRT error circle of IGRW\,151019 (see section \ref{sec:IGRW151019results}).}
\end{itemize}

Table \ref{tab:uvotobs} contains the magnitudes and limits that were obtained for each source in each available filter. Some sources occurred in crowded fields affecting background subtraction. We also included Galactic reddening values for each source. The $A_{V}$ values were taken from the Infrared Science Archive (IRSA)\footnote{\label{IRSA} http://irsa.ipac.caltech.edu/applications/DUST/} using the method described in \cite{Schlafly2011}.

\begin{table*}
\centering
	\begin{tabular}{|c|c|c|c|c|c|c|c|c|c|}
	\hline 
Name & \multicolumn{1}{c}{white} & \multicolumn{1}{c}{v} & \multicolumn{1}{c}{b} & \multicolumn{1}{c}{u} & \multicolumn{1}{c}{w1} & \multicolumn{1}{c}{m2} & \multicolumn{1}{c}{w2} & Source? & $A_{V}$ (Mag) 
	\\ \hline
    IGRW\,151019 & - & - & - & - & $>21.75$ & 22.16($\pm0.15$) & - & Yes & 0.75 \\ \hline
    GRB\,150831A & $>21.54$ & $>19.82$ & $>19.90$ & $>21.34$ & $>21.85$ & $>22.39$ & $>23.10$ & No & 0.30 \\ \hline
    GRB\,150305A & - & - & - & - & $>22.68$ & $>22.74$ & - & No & 0.45 \\ \hline
	GRB\,121212A & 23.89($\pm0.38$) & $>20.22$ & $>20.75$ & $>22.07$ & $>23.00$ & $>22.73$ & $>22.95$ & No & 0.18 \\ \hline
	GRB\,100909A & $>22.52$ & $>20.07$ & $>21.26$ & $>21.78$ & $>21.73$ & $>22.19$ & $>21.99$ & No & 1.37 \\ \hline
	GRB\,091111 & - & 19.48($\pm0.24$)$^\textbf{*}$ & 18.92($\pm0.28$)$^\textbf{*}$ & 21.76($\pm0.23$)$^\textbf{*}$ & $>21.94$ & $>22.35$ & $>22.32$ & No & 4.79 \\ \hline
	\end{tabular}
	\caption[UVOT Table]{UVOT multi-band magnitudes (AB) and $3\sigma$ upper limits of the 6 XRT detected sources. Filters ordered with decreasing wavelength.

$^\textbf{*}$The GRB\,091111 XRT position was within 10 arcsec of a very bright, saturated source and further analysis of the images suggest that the detections in the v, b and u filters are probably not real.}
	\label{tab:uvotobs}
\end{table*} 

\subsection{GRB\,150305A} \label{sec:GRB150305Aresults}
\begin{figure*}
\includegraphics[width=\textwidth]{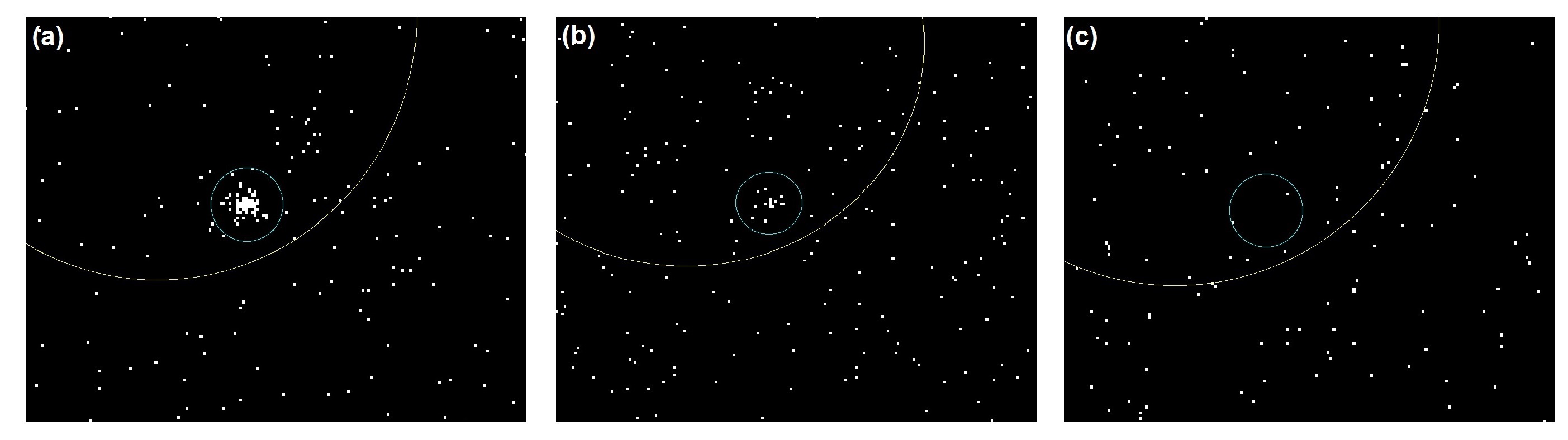}
\caption{Images showing the X-ray source IGRW\,150305/GRB\,150305A (cyan) within the \textit{INTEGRAL} error circle (yellow). The images correspond to observation times of $(1.7-2.7)\times 10^{4}$ s (a), $(1.2-1.3)\times 10^{5}$ s (b) and $(7.0-7.4)\times 10^{5}$ s (c) after the GRB occurred with XRT exposure times of 3.0 ks (a), 3.7 ks (b) and 2.4 ks (c) respectively. The point source clearly fades over time and is undetectable after $\approx 7.4\times 10^{5}$ s ($8-9$ days).}
\label{fig:150305A}
\end{figure*}

IGRW\,150305 was confirmed to have a fading X-ray afterglow after requesting 3 ToOs over a time period of $\approx 8-9$ days, the first of which began 17 ks after the WEAK trigger (see figure \ref{fig:150305A}). A marginal detection was made in the white UVOT filter. A Vizier search of the GRB position revealed no optical or X-ray catalogue matches for the XRT and UVOT positions.

The light curve of GRB\,150305A was poorly sampled due to the limited exposure from the ToOs but it is consistent with a decay slope of ${\rm \alpha} \approx 1$. An optimized fit could not be produced for the light curve so this decay slope is a rough estimation. Obtaining a spectrum is not possible due to the low number of counts detected: 102 in 6620 s. This was a detection of a new GRB directly from \textit{Swift} follow-up of a WEAK trigger and was not identified elsewhere \citep{Starling2015}.

\subsection{IGRW\,151019 - Active Galactic Nucleus candidate} \label{sec:IGRW151019results}
IGRW\,151019 had showed no signs of fading after 4 weeks; initially the count rate was $4.0(\pm 2.7)\times 10^{-3}$ s$^{-1}$ increasing by a factor of $\approx 1-8$ in the X-ray band over a time period of $\approx 1$ month. There were only two observations of this source as a third ToO was not required as the source clearly was not fading. IGRW\,151019 may therefore not be a GRB but a steady source. Analysis of the spectrum when fit with an absorbed power law gave a total column density, $ N_{\rm{H}} = 4.0^{+5.0}_{-3.0}\times 10^{21}$ cm$^{-2}$ and a photon index, $\Gamma = 1.71\pm 0.37$, which is broadly consistent with that of an AGN \citep{Nandra1994,Tozzi2006,Brightman2011}.

Inside the \textit{Swift} XRT error circle lies the catalogued AllWISE source J193108.05+310756.4 \citep{Cutri2014}. The source is within 1.8 arcsec of the centre of the XRT position and within the 90 per cent XRT error region. The UVOT source we detected (table \ref{tab:uvotobs})  for IGRW\,151019 in the m2 filter has is RA, Dec (J2000) 292.78334, 31.13252 deg with a $1\sigma$ positional error of 0.49 arcsec. A catalogued Gaia source has a position coincident to the UVOT source to within 0.5 arcsec. However, these sources lie just outside the 90 per cent XRT error region so we cannot confirm their association with the new X-ray transient. In addition, the Galactic extinction in this direction, $A_{V} \approx 0.75$. Examining the WISE source in more depth we find that its WISE colours, ${\rm W1-W2}=0.8$ and ${\rm W2-W3}=2.4$, are consistent with that of an AGN \citep{Mingo2016}. It is possible that a transient event may have caused the initial \textit{INTEGRAL} WEAK trigger and the steady source may simply be a chance coincidence observation. However, this is unlikely and we conclude that IGRW\,151019 is likely an AGN.

\section{IBAS and Swift GRB sample properties} \label{sec:IBASsamps}
In section \ref{sec:intro} we discussed that a low-luminosity GRB population could exist and that \textit{INTEGRAL} may be capable of detecting it. Including the WEAK alert GRBs we have confirmed, the IBAS GRB sample size currently stands at 114. The \textit{Swift} sample size stands at 1060 GRBs with XRT detections for 846 GRBs (all values correct as of 2016 July 1 and \textit{Swift} numbers were taken from the \textit{Swift} GRB Table\footnote{\label{swifttable} ${\rm http://swift.gsfc.nasa.gov/archive/grb\_table/}$}).

\begin{figure}
\includegraphics[width=\linewidth]{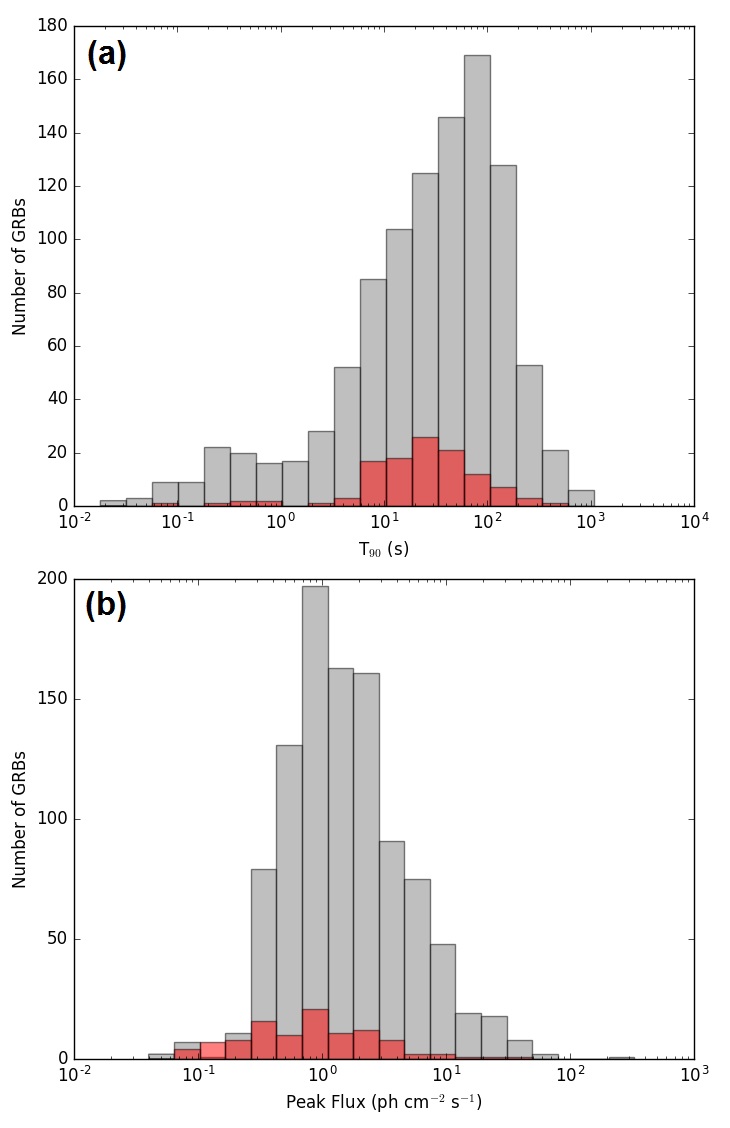}
\caption{Histograms of T$_{90}$ and peak flux distributions of both the \textit{Swift} BAT (light grey) and IBAS (red) GRB samples. The peak flux values are measured between $15-150$ keV for \textit{Swift} and $20-200$ keV for \textit{INTEGRAL}. The fraction of short GRBs in the \textit{Swift} and IBAS samples are 9.6 per cent and 5.3 per cent respectively.}
\label{fig:svidists} 
\end{figure}

Figure \ref{fig:svidists} shows the T$_{90}$ and peak flux distributions of the \textit{INTEGRAL} IBAS and \textit{Swift} BAT samples. Analysis shows that the IBAS sample has a lower mean T$_{90}$ (47 s compared to 70 s). However, the \textit{Swift} sample has a higher percentage of short GRBs compared to IBAS; 95 short GRBs out of the 992 \textit{Swift} GRBs with measured T$_{90}$ (9.6 per cent) compared to 6 short GRBs out of 114 \textit{INTEGRAL} GRBs (5.3 per cent). The mean peak flux of the IBAS GRB sample is also lower than that of the \textit{Swift} sample (2.0 ph cm$^{-2}$ s$^{-1}$ compared to 3.6 ph cm$^{-2}$ s$^{-1}$) meaning \textit{INTEGRAL} routinely reaches lower peak flux values as a proportion of the total sample. With a lower average T$_{90}$ and peak flux it is more likely that the fluence distribution of \textit{INTEGRAL} may be skewed towards fainter GRBs than \textit{Swift} and with the addition of the lower IBIS sensitivity (discussed in section \ref{sec:intro}) \textit{INTEGRAL} may be better suited to probing this lower luminosity GRB population.

\subsection{Can IBAS be used to probe low fluence GRBs?} \label{sec:IBASprops}
Our IBAS GRB sample contains 92 of the 114 IBAS GRBs, i.e. all those that have calculated fluence values. All IBAS T${\rm _{90}}$ values were taken from the IBAS webpage\textsuperscript{\ref{IBAS}} but did not have any associated error limits. The IBAS GRB fluxes were calculated in {\sc xspec} using a simple power law/cut-off power law model. The flux values were then multiplied by the T${\rm_{90}}$ values to calculate the fluences. Several fluence values were taken from \cite{Vianello2009} and \cite{Bosnjak2014} and the properties of all IBAS GRBs with published and estimated measurements can be found in table \ref{tab:ibasgrbs}.

\begin{figure}
\includegraphics[width=\linewidth]{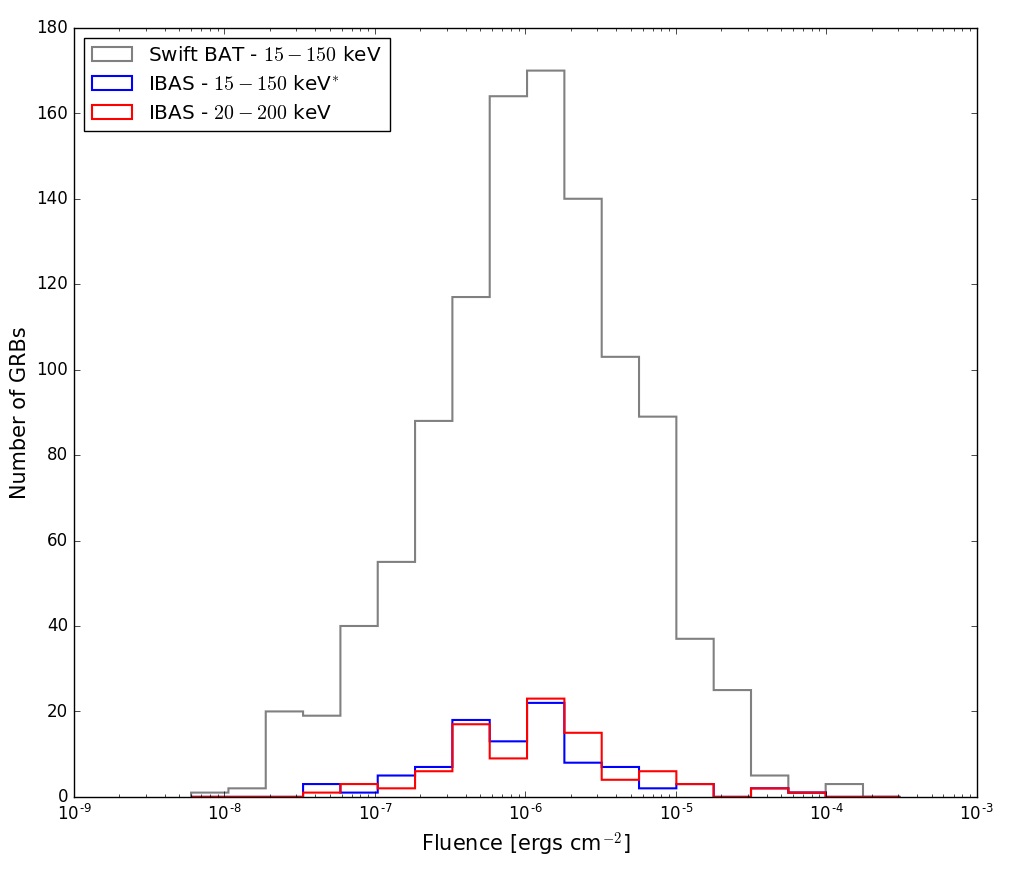}
\caption{Fluence distributions of both the \textit{Swift} BAT and the IBAS GRB samples. 
$^{\textbf{*}}$Calculated using method described in section \ref{sec:IBASprops}.} 
\label{fig:fluencehist}  
\end{figure}

For the IBAS GRBs detected by the XRT, we calculated the X-ray flux values at 11 hours by fitting a series of single/broken power laws to the \textit{Swift} X-ray afterglow light curves and performing an f-test to determine the best-fitting model. Once the best fitting model was obtained this was extrapolated to 39600s (11 hours) and an estimate for the X-ray flux was determined. For the 1026 \textit{Swift} GRBs with fluences, both the fluence, associated errors (at 90 per cent confidence) and X-ray flux at 11 hours values were taken from the NASA Goddard Space Flight Center \textit{Swift} GRB Table\textsuperscript{\ref{IBAS}}. X-ray flux values < $10^{-14}$ erg cm$^{-2}$ s$^{-1}$ were omitted as they were deemed too faint for \textit{Swift} to detect and are therefore non-physical measurements. This resulted in 824 \textit{Swift} GRBs and 33 IBAS GRBs with calculated X-ray flux values. It must be noted that the \textit{Swift} GRB fluences are measured in the energy range of $15-150$ keV whereas the \textit{INTEGRAL} GRB fluences are measured between $20-200$ keV. We found that the flux, and therefore, fluence ratios between these two energy bands was ${\rm \frac{f_{20-200}}{f_{15-150}}}\approx 1.22$ when measured from spectral fits for a small number of typical sources from our sample. We highlight that this is a mean ratio used to give an indication of the IBAS fluence values in the BAT energy band and will vary between GRBs within the sample. We also note that the T${\rm _{90}}$ can vary between different energy bands but we assumed that it remains constant for this conversion. Figure \ref{fig:fluencehist} shows the \textit{Swift} BAT GRB fluence distribution overlaid with the IBAS GRB fluence distribution both in the $20-200$ keV and converted $15-150$ keV energy bands.

To determine if the IBAS and \textit{Swift} GRB sample fluence distributions came from the same underlying population, a K-S test was performed on the \textit{Swift} and IBAS fluence values in their respective $15-150$ and $20-200$ keV energy bands. If the test rejected the null hypothesis; that the underlying distribution of the samples was the same at a 95 per cent confidence level (p $<0.05$) then it was assumed that the samples were not part of the same underlying distribution. The K-S test returned a $p$-value of 0.37 for the fluence distribution so we cannot reject the null hypothesis and we conclude that the \textit{INTEGRAL} IBAS and \textit{Swift} GRB samples most likely belong to the same distribution. Moreover, the mean fluence values are similar, $3.66\times 10^{-6}$ erg cm$^{-2}$ and $3.94\times 10^{-6}$ erg cm$^{-2}$ for the \textit{Swift} and IBAS samples respectively. Converting the IBAS fluence values into the $15-150$ keV band using the ratio calculated previously gives a mean fluence value of $3.23\times 10^{-6}$ erg cm$^{-2}$ and when compared to the \textit{Swift} distribution gives a K-S $p$-value of 0.06. This results in the same conclusion as before; the two distributions most probably belong to the same fluence distribution.

We also analysed the correlations between GRB fluence, T$_{90}$ and X-ray flux at 11 hours for both the IBAS and \textit{Swift} GRB samples. The Spearman rank coefficients for fluence - T$_{90}$ were $0.52(\pm0.07)$ and $0.66(\pm0.02)$ corresponding to $p$-values of $5.7\times 10^{-8}$ and $3.8\times 10^{-130}$ for the IBAS and \textit{Swift} samples respectively. The Spearman rank coefficients for fluence - X-ray flux were $0.65(\pm0.11)$ and $0.61(\pm0.02)$ corresponding to $p$-values of $2.4\times 10^{-5}$ and $3.96\times 10^{-90}$ for the IBAS and \textit{Swift} samples respectively. These values show these parameters exhibit significant correlation. Similar correlations have been reported in previous investigations \citep{Gehrels2008,Evans2009,Margutti2013,Grupe2013}. These authors acknowledge a wide spread in the data (to within an order of magnitude) due to a range of factors. We do not have available errors for the IBAS T$_{90}$ values and these can be underestimated for very long GRBs. Our extrapolation values of X-ray flux at 11 hours do not have any associated errors and for some GRBs can only be taken as rough estimates due to the low number of data bins. Additionally we have used the observed X-ray flux, prior to correction for line-of-sight absorption. Although we have not fully accounted for these effects our correlations are significant. We conclude that our correlations agree with similar correlations from previous investigations.   

\begin{figure*}
\includegraphics[width=\textwidth]{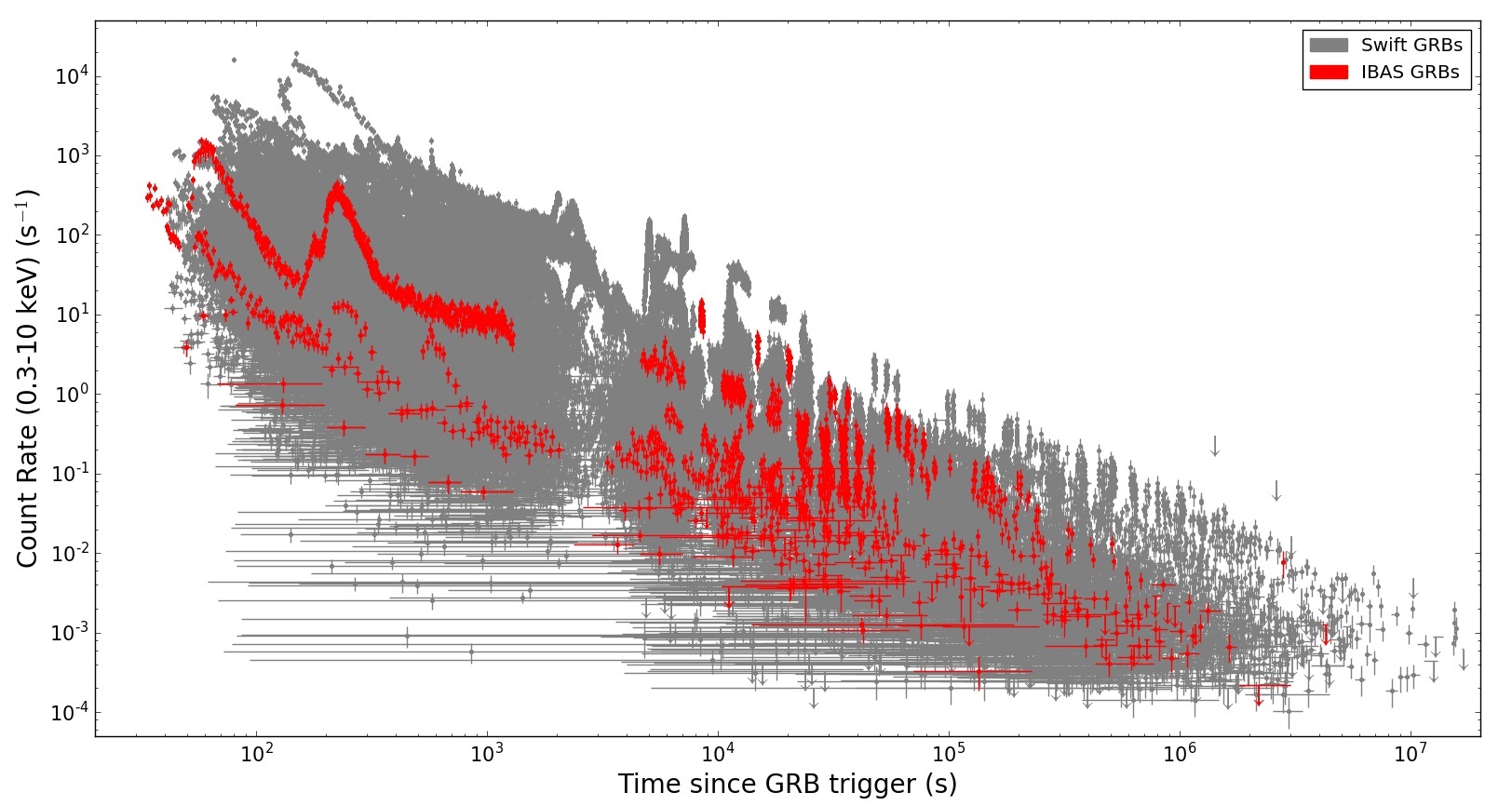}
\caption{X-ray afterglows of the \textit{Swift} GRB sample and the 54 IBAS GRBs observed by \textit{Swift}/XRT.} 
\label{fig:xrtplot}  
\end{figure*}

Figure \ref{fig:xrtplot} shows the X-ray afterglows of the IBAS and \textit{Swift} GRB samples. The plot highlights that the X-ray afterglow distribution of the IBAS sample sits comfortably within the \textit{Swift} X-ray GRB afterglow distribution. The mean X-ray flux values at 11 hours for the IBAS and \textit{Swift} samples are $2.85\times 10^{-12}$ erg cm$^{-2}$ s$^{-1}$ and $1.48\times 10^{-12}$ erg cm$^{-2}$ s$^{-1}$ showing that on average, the X-ray flux of the \textit{Swift} sample GRBs is lower. However, only 54 IBAS GRBs were detected by the XRT; not all were followed up, some were non-detections, and only 33 were sufficiently sampled to obtain a value of X-ray flux at 11 hours.

\textit{Swift} and \textit{INTEGRAL} regularly detect similar fluence GRBs, however, the \textit{Swift} sample has a low fluence, short GRB tail that the IBAS sample does not. \textit{Swift} has also detected six GRBs which may belong to a further subclass of "ultra-long" GRBs where T${\rm _{90}}$ values $\gtrsim 1000$ s \citep{Gendre2013,Virgili2013,Evans2014,Levan2014,Cucchiara2015}. However, with this low number of "ultra-long" GRBs ($<1\%$ of the sample) we do not expect to have detected any with IBAS. \textit{Swift} has detected $\approx 10$ times the number of GRBs than IBAS has detected. With such a large \textit{Swift} sample you would expect to see some very faint and very long GRBs and the differences in the distributions may arise from the smaller IBAS sample size and low number statistics. From this investigation we conclude that the \textit{Swift} and IBAS GRB distributions are similar but not identical.   

\section{Conclusions} \label{sec:concl}
We investigated 15 \textit{INTEGRAL} WEAK triggers utilising \textit{Swift} for follow-up observations. Among these WEAK triggers, we confirm seven astrophysical events - six GRBs and one candidate AGN. IGRW\,150305 found directly from one of our chosen ToOs was identified as a GRB through this ToO campaign alone. 

Comparisons of the fluence distributions of the full IBAS and \textit{Swift} GRB samples showed that the two are similar but not identical. We also confirm correlations between the gamma-ray and X-ray properties found in previous investigations for both samples. Both the IBAS GRB fluence and X-ray afterglow light curve distributions comfortably lie within the \textit{Swift} distributions. We conclude that \textit{Swift} and IBAS typically reach similar fluence limits,  while \textit{Swift} appears to be more sensitive to short, low fluence GRBs. 

We only sample $\approx 4$ per cent of the total WEAK trigger population. Hence we do not make any statistical statements for the total sample. We have shown that \textit{INTEGRAL} can detect real GRB events below the STRONG threshold, along with other high energy transients and variables such as AGN. This allows future work to uncover the nature of yet more WEAK triggers to determine whether \textit{INTEGRAL} can detect fainter GRBs.

\section{Acknowledgements}
The research leading to these results has received funding from the European Union's Horizon 2020 Programme under AHEAD project (grant agreement n. 654215). A. B. Higgins is supported by an Science and Technology Facilities Council (STFC) studentship. R. L. C. Starling and K. Wiersema acknowledge support from STFC. This work made use of data supplied by the UK \textit{Swift} Science Data Centre at the University of Leicester which is supported by the UK Space Agency. D. Gotz acknowledges the financial support of the UnivEarthS Labex program at Sorbonne Paris Cit\'e (ANR-10-LABX-0023 and ANR-11-IDEX-0005-02). S. Mereghetti acknowledges the support of ASI/INAF agreement No. 2016-025-R.0. This research has made use of the VizieR catalogue access tool, CDS, Strasbourg, France. We also thank the referee for useful comments and feedback.

During the reviewing process for this paper, our colleague Neil Gehrels passed away. We would like to acknowledge not only his input into this work, but his immense contributions to high energy astrophysics as a whole.

\bibliographystyle{mnras}
\bibliography{adambib}

\appendix
\section{Supplementary Tables} \label{sec:appendixA}
Supplementary material containing IBAS GRB gamma-ray and x-ray data used during this investigation.
 
\begin{table*}
\centering
	\begin{tabular}{|c|c|c|c|}
    \hline
  \multicolumn{1}{|p{2.5cm}|}{\centering Name}
& \multicolumn{1}{|p{3.5cm}|}{\centering Fluence [$20-200$ keV] ($10^{-7}$ erg cm$^{-2}$)} 
& \multicolumn{1}{|p{3.0cm}|}{\centering X-ray Flux at 11 hours [$0.3-10$ keV] ($10^{-12}$ erg cm$^{-2}$ s$^{-1}$)}
& \multicolumn{1}{|p{1.0cm}|}{T$_{{\rm 90}}$ (s)}
  \\ \hline
GRB\,030227$^{\textbf{*}}$ & $6.10^{+3.50}_{-5.90}$ & - & 15 \\ \hline
GRB\,030320$^{\textbf{*}}$ & $54.2^{+13.3}_{-11.7}$ & - & 48 \\ \hline
GRB\,030501$^{\textbf{*}}$ & $17.2^{+1.60}_{-3.10}$ & - & 25 \\ \hline
GRB\,030529$^{\textbf{\#}}$ & 0.52 & - & 16 \\ \hline
GRB\,031203$^{\textbf{*}}$ & $10.6^{+2.70}_{-3.00}$ & - & 19 \\ \hline
GRB\,040106$^{\textbf{*}}$ & $95.0^{+23.0}_{-30.0}$ & - & 47 \\ \hline 
GRB\,040223$^{\textbf{*}}$ & $27.2^{+0.80}_{-1.90}$ & - & 258 \\ \hline
GRB\,040323$^{\textbf{*}}$ & $20.6^{+2.30}_{-2.90}$ & - & 14 \\ \hline
GRB\,040403$^{\textbf{*}}$ & $4.00^{+1.60}_{-3.70}$ & - & 15 \\ \hline
GRB\,040422$^{\textbf{*}}$ & $4.90^{+1.00}_{-3.60}$ & - & 10 \\ \hline
GRB\,040624$^{\textbf{\#}}$ & 4.81 & - & 27 \\ \hline
GRB\,040730$^{\textbf{*}}$ & $6.30^{+4.40}_{-3.30}$ & - & 42 \\ \hline
GRB\,040812$^{\textbf{\#}}$ & 1.40 & - & 8 \\ \hline
GRB\,040827$^{\textbf{*}}$ & $11.1^{+2.80}_{-4.00}$ & - & 32 \\ \hline
GRB\,040903$^{\textbf{\#}}$ & 0.96 & - & 7 \\ \hline
GRB\,041015$^{\textbf{\#}}$ & 5.12 & - & 30 \\ \hline
GRB\,041218$^{\textbf{*}}$ & $58.2^{+3.50}_{-3.70}$ & - & 38 \\ \hline
GRB\,041219A$^{\textbf{*}}$ & $867^{+0.50}_{-129}$ & - & 239 \\ \hline
GRB\,050129$^{\textbf{\#}}$ & 4.10 & - & 30 \\ \hline
GRB\,050223 & $10.8^{+2.70}_{-2.10}$ & 0.19 & 30 \\ \hline
GRB\,050502A$^{\textbf{*}}$ & $13.9^{+1.10}_{-4.00}$ & - & $>11$ \\ \hline
GRB\,050504$^{\textbf{*}}$ & $10.0^{+4.10}_{-4.50}$ & - & 44 \\ \hline
GRB\,050520$^{\textbf{*}}$ & $16.6^{+4.90}_{-5.00}$ & 0.20 & 52 \\ \hline
GRB\,050522$^{\textbf{\#}}$ & 0.69 & - & 11 \\ \hline
GRB\,050525A$^{\textbf{*}}$ & $154^{+5.70}_{-8.40}$ & 1.5 & 9 \\ \hline
GRB\,050626$^{\textbf{*}}$ & $6.30^{+0.40}_{-1.00}$ & - & 52 \\ \hline
GRB\,050714A & $5.58^{+2.75}_{-1.84}$ & - & 34 \\ \hline
GRB\,050918$^{\textbf{*}}$ & $30.2^{+10.5}_{-9.0}$ & - & 280 \\ \hline
GRB\,050922A$^{\textbf{\#}}$ & 0.59 & - & 10 \\ \hline
GRB\,051105B$^{\textbf{*}}$ & $2.80^{+1.50}_{-2.00}$ & - & 14 \\ \hline
GRB\,051211B$^{\textbf{*}}$ & $16.1^{+4.60}_{-3.30}$ & 0.92 & 47 \\ \hline
GRB\,060114$^{\textbf{*}}$ & $16.0^{+4.60}_{-3.30}$ & - & 80 \\ \hline
GRB\,060130$^{\textbf{\#}}$ & 2.25 & - & 19 \\ \hline
GRB\,060204A$^{\textbf{*}}$ & $4.80^{+2.40}_{-3.30}$ & - & 52 \\ \hline
GRB\,060428C$^{\textbf{*}}$ & $18.6^{+2.20}_{-3.90}$ & - & 10 \\ \hline
GRB\,060901$^{\textbf{*}}$ & $62.2^{+3.50}_{-5.90}$ & 1.2 & 16 \\ \hline
GRB\,060930$^{\textbf{\#}}$ & 2.63 & - & 9 \\ \hline
GRB\,060912B$^{\textbf{*}}$ & $12.0^{+5.80}_{-5.10}$ & - & 140 \\ \hline
GRB\,061025$^{\textbf{*}}$ & $10.1^{+1.30}_{-4.80}$ & 0.14 & 11 \\ \hline
	\end{tabular}
\end{table*}

\begin{table*}
\centering
	\begin{tabular}{|c|c|c|c|}
    \hline
  \multicolumn{1}{|p{2.5cm}|}{\centering Name}
& \multicolumn{1}{|p{3.5cm}|}{\centering Fluence [$20-200$ keV] ($10^{-7}$ erg cm$^{-2}$)} 
& \multicolumn{1}{|p{3.0cm}|}{\centering X-ray Flux at 11 hours [$0.3-10$ keV] ($10^{-12}$ erg cm$^{-2}$ s$^{-1}$)}
& \multicolumn{1}{|p{1.0cm}|}{T$_{{\rm 90}}$ (s)}
  \\ \hline
GRB\,061122$^{\textbf{*}}$ & $155^{+3.40}_{-5.30}$ & 2.2 & 12 \\ \hline
GRB\,070309 & $4.93^{+3.12}_{-1.98}$ & - & 22 \\ \hline
GRB\,070311$^{\textbf{*}}$ & $23.6^{+1.70}_{-5.30}$ & 1.22 & 32 \\ \hline
GRB\,070615 & 2.01 & - & 15 \\ \hline
GRB\,070707 & $3.58^{+4.04}_{-1.94}$ & - & 0.7 \\ \hline
GRB\,070925$^{\textbf{*}}$ & $36.1^{+1.70}_{-3.40}$ & - & 19 \\ \hline
GRB\,071003 & $94.6^{+4.22}_{-2.96}$ & 3.5 & 38 \\ \hline
GRB\,071109$^{\textbf{*}}$ & $3.60^{+4.00}_{-3.50}$ & - & 30 \\ \hline
GRB\,080120 & $13.2^{+17.0}_{-7.67}$ & 0.13 & 15 \\ \hline
GRB\,080603A & $12.3^{+1.70}_{-5.90}$ & 1.5 & 150 \\ \hline
GRB\,080613A$^{\textbf{*}}$ & $12.3^{+1.70}_{-5.90}$ & - & 30 \\ \hline
GRB\,080723B$^{\textbf{*}}$ & $396^{+6.70}_{-6.70}$ & 12.6 & 95 \\ \hline
GRB\,080922$^{\textbf{*}}$ & $17.3^{+6.90}_{-6.50}$ & - & 60 \\ \hline
GRB\,081003B$^{\textbf{*}}$ & $26.2^{+2.00}_{-24.5}$ & - & 20 \\ \hline
GRB\,081016$^{\textbf{*}}$ & $22.0^{+1.40}_{-4.50}$ & - & 30 \\ \hline
GRB\,081204$^{\textbf{*}}$ & $5.10^{+5.10}_{-4.80}$ & - & 12 \\ \hline
GRB\,090107B$^{\textbf{*}}$ & $12.4^{+1.30}_{-4.60}$ & 0.73 & 15 \\ \hline
GRB\,090625B$^{\textbf{*}}$ & $12.4^{+1.20}_{-2.00}$ & 0.38 & 8 \\ \hline
GRB\,090702 & $1.93^{+1.44}_{-0.81}$ & - & 6 \\ \hline
GRB\,090704$^{\textbf{*}}$ & $54.0^{+4.90}_{-8.00}$ & - & 70 \\ \hline
GRB\,090814B$^{\textbf{*}}$ & $15.1^{+2.30}_{-2.40}$ & 1.4 & 42 \\ \hline
GRB\,090817$^{\textbf{*}}$ & $18.7^{+10.9}_{-9.80}$ & 2.4 & 30 \\ \hline
GRB\,091111 & $20.0^{+5.90}_{-0.82}$ & - & 100 \\ \hline
GRB\,091202 & $7.03^{+3.02}_{-2.35}$ & - & 25 \\ \hline
GRB\,091230 & $17.9^{+20.5}_{-9.57}$ & - & 70 \\ \hline
GRB\,100103A$^{\textbf{*}}$ & $52.5^{+2.10}_{-4.00}$ & 2.1 & 30 \\ \hline
GRB\,100518A$^{\textbf{*}}$ & $5.20^{+4.40}_{-3.80}$ & 0.87 & 25 \\ \hline
GRB\,100713A & $5.65^{+2.65}_{-1.80}$ & 0.20 & 20 \\ \hline
GRB\,100909A & $21.5^{+7.00}_{-4.70}$ & 0.26 & 60 \\ \hline
GRB\,101112A$^{\textbf{*}}$ & $21.1^{+4.40}_{-7.40}$ & 0.50 & 6 \\ \hline
GRB\,110206A & $17.2^{+11.6}_{-6.10}$ & 2.0 & 15 \\ \hline
GRB\,110708A$^{\textbf{*}}$ & $24.8^{+1.90}_{-4.60}$ & - & 50 \\ \hline
GRB\,110903A$^{\textbf{*}}$ & $148^{+11.9}_{-17.5}$ & 3.8 & 430 \\ \hline
GRB\,120202A & $8.00^{+2.10}_{-7.70}$ & - & 70 \\ \hline
GRB\,120419A & $3.88^{+6.18}_{-2.49}$ & - & 15 \\ \hline
GRB\,120711A & $440^{+50.0}_{-5.00}$ & 40 & 135 \\ \hline
GRB\,121102A & $24.1^{+12.4}_{-8.10}$ & 0.56 & 25 \\ \hline
GRB\,121212A & 1.50 & 0.46 & 10 \\ \hline
	\end{tabular}
\end{table*}

\begin{table*}
\centering
	\begin{tabular}{|c|c|c|c|}
    \hline
  \multicolumn{1}{|p{2.5cm}|}{\centering Name}
& \multicolumn{1}{|p{3.5cm}|}{\centering Fluence [$20-200$ keV] ($10^{-7}$ erg cm$^{-2}$)} 
& \multicolumn{1}{|p{3.0cm}|}{\centering X-ray Flux at 11 hours [$0.3-10$ keV] ($10^{-12}$ erg cm$^{-2}$ s$^{-1}$)}
& \multicolumn{1}{|p{1.0cm}|}{T$_{{\rm 90}}$ (s)}
  \\ \hline
GRB\,130513A & $17.0^{+10.3}_{-6.50}$ & - & 50 \\ \hline
GRB\,130514B & $10.2^{+14.4}_{-6.20}$ & 1.6 & 10 \\ \hline
GRB\,130903A & $17.1^{+8.10}_{-5.40}$ & - & 70 \\ \hline
GRB\,131122A & $24.8^{+12.3}_{-8.20}$ & - & 80 \\ \hline
GRB\,140206A & $16.0^{+3.00}_{-3.00}$ & 9.2 & ${\rm >60}$ \\ \hline
GRB\,140320B & $12.7^{+11.8}_{-5.94}$ & 0.55 & 100 \\ \hline
GRB\,140320C & 3.52 & - & 30 \\ \hline
GRB\,140815A & $5.00^{+5.10}_{-2.59}$ & - & 8 \\ \hline
GRB\,141004A & $6.92^{+6.88}_{-3.40}$ & 0.09 & 4 \\ \hline
GRB\,150219A & $57.1^{+14.9}_{-11.2}$ & 0.62 & 60 \\ \hline
GRB\,150305A & $12.1^{+14.2}_{-6.45}$ & - & 100 \\ \hline
GRB\,150831A & $\approx 3$ & 0.33 & 2 \\ \hline
GRB\,151120A & $\approx 20$ & 0.66 & 50 \\ \hline
GRB\,160221A & $\approx 5$ & - & 10 \\ \hline
GRB\,160629A & $\approx 60$ & - & 100 \\ \hline
	\end{tabular}
	\caption[IBAS XRT Detected GRBs]{The prompt emission fluence, T$_{90}$ and X-ray afterglow flux of all 92 IBAS GRBs within our sample. All T$_{90}$ values were taken from IBAS\textsuperscript{\ref{IBAS}}. Some GRBs were not observed by \textit{Swift} or had very poorly sampled \textit{Swift} XRT afterglows and therefore an X-ray flux at 11 hours could not be obtained. Some fluence values contain no errors as the fitted model would not converge and would not provide an error on the normalisation. This meant no error could be found on the flux values and therefore the fluence values but the values should still be representative of the actual fluence. Fluence values for GRB\,150831, GRB\,151120A, GRB\,160221A, GRB\,160629A are approximations from the GCN Circulars \citep{Mereghetti2015a,Mereghetti2015b,Mereghetti2016,Gotz2016} as the spectral data were not yet public.

$^{\textbf{*}}$ Fluence values taken from \cite{Bosnjak2014}.

$^{\textbf{\#}}$ Fluence values taken from \cite{Vianello2009}.}
	\label{tab:ibasgrbs}
\end{table*}

\label{lastpage}

\end{document}